\documentclass[preprintnumbers,amsmath,amssymb]{revtex4}

\usepackage{graphicx}
\usepackage{dcolumn}
\usepackage{bm}

\begin{document}
\draft
\title{Self-organization in dissipative optical lattices}
\author{G. Baris Ba\u{g}c{\i} and Ugur Tirnakli}
\email{ugur.tirnakli@ege.edu.tr}
\address {Department of Physics, Faculty of Science, Ege University, 35100 Izmir, Turkey}

\pagenumbering{arabic}

\begin{abstract}

We show that the transition from Gaussian to the $q$-Gaussian
distributions occurring in atomic transport in dissipative optical
lattices can be interpreted as self-organization by recourse to a
modified version of Klimontovich's S-theorem. As a result, we find
that self-organization is possible in the transition regime, only
where the second moment $\left\langle p^{2}\right\rangle $ is
finite. Therefore, the nonadditivity parameter $q$ is confined
within the range $1<q<\frac{5}{3}$, although whole spectrum of $q$
values i.e., $1<q<3$, is considered theoretically possible. The
range of $q$ values obtained from the modified S-theorem is also
confirmed by the experiments carried out by Douglas \textit{et al.}
[Phys. Rev. Lett. 96, 110601 (2006)].

\end{abstract}

\newpage \setcounter{page}{1}

\maketitle

\section{\protect\bigskip Introduction}

The importance of understanding physical phenomena exhibiting
asymptotical inverse power law distributions becomes apparent if we
consider its emergence in such diverse fields as subregion laser
cooling [1], the heartbeat histograms of healthy patients [2],
blinking nanocrystals [3], rheology of steady-state draining foams
[4], econophysics [5], earthquake models [6], to name but a few.

One such distribution is the Tsallis distribution [7], which found
many applications in different fields [8]. Recently, it was pointed
out that the momentum distribution of cold atoms in a dissipative
optical lattice is a Tsallis distribution [9]. Soon, this
theoretical prediction has been verified experimentally [10].

Three different regimes of atomic transport is possible in an
optical lattice (see Ref. [11] for details), depending on the depth
of the optical potential. In the intermediate regime between the
diffusive motion in deep potentials and ballistic motion in shallow
potentials, the atomic transport is governed by the Tsallis
distribution [9, 10]. In fact, it has been shown, both theoretically
and experimentally, that a transition from an ordinary Gaussian to
the $q$-Gaussian distribution occurs by changing the ratio of the
recoil energy to the potential depth [9, 10].

Our aim in this paper is to show that this transition can be
interpreted as a self-organization in the context of modified
S-theorem. Klimontovich [12-17] originally proposed S-theorem
(Self-organization theorem) in order to generalize Gibbs' theorem
[18] to open systems, since the latter rests on the assumption that
all the compared distributions have the same mean energy values.
However, this assumption does not hold when one studies open systems
where the mean energy of the system is not constant due to the
interaction of the system with the environment (e.g., a laser field in 
the case of optical lattices).
Therefore, Klimontovich defined a quantity called effective energy.
Equating the mean effective energies of different states and then
comparing the associated entropies, he showed that S-theorem orders
the associated entropies in such a way that the state closer to
equilibrium possesses a greater entropy compared to the other
states. In other words, we have a more ordered state as the control
parameter increases. This decrease of entropy on ordering is called
self-organization by Haken [19]. We also mention that the process of
equating mean effective energies of different states is called
renormalization by Klimontovich. As a result, S-theorem did not only
succeed in generalizing Gibbs' theorem but gained recognition as a
criterion of self-organization in open systems, since it allows us
to compare even the stationary nonequilibrium distributions [12-17].

We also mention that the use of S-theorem is not limited to
analytical models. It has also been used for many numerical models
such as logistic map [20], heart rate variability [21, 22] and the
analysis of electroencephalograms of epilepsy patients [23] as a
criterion of self-organization.

However, S-theorem rests on the use of Boltzmann-Gibbs (BG) entropy.
Therefore, it is useful to analyze transition regimes where the
stationary distributions are of exponential form. On the other hand,
the study of open systems may result in stationary distributions of
the inverse power law form. Motivated by this fact, a generalization
of the S-theorem has recently been provided by the employment of
nonadditive Tsallis entropy [24, 25]. This generalized nonadditive
S-theorem can be used to analyze the transition regimes in open
systems, where the stationary distributions are $q$-exponentials,
which asymptotically yield to inverse power law distributions.

Although the dissipative optical lattices serve as open systems due
to their interaction with laser fields, neither original S-theorem
due to Klimontovich nor its nonadditive generalization can be used
for the description of self-organization in the aforementioned
intermediate region of atomic transport. The reason is that this
region exhibits a transition from Gaussian to $q$-Gaussian
distributions. Therefore, the intermediate region requires an
intermediate treatment between the two formulations of the
S-theorem. This intermediate S-theorem, called modified S-theorem,
is provided here for the explicit purpose of applying it to the
dissipative optical lattices.

The paper is organized as follows: In Section II, we review the
original (i.e., additive) and nonadditive S-theorems for open
systems. The modified S-theorem and its immediate application to
dissipative open lattices are presented in Section III. Concluding
remarks will be presented in Section IV.

\section{The additive and nonadditive S-theorems}

The additive S-theorem due to Klimontovich [12-17] requires the use
of two distinct probability distributions, i.e., $r_{eq}$ and $p$,
corresponding to equilibrium and nonequilibrium states,
respectively. The stationary equilibrium distribution is defined as
the distribution corresponding to the state where the relevant
control parameter is set to zero. Similarly, any other stationary
state with non-zero control parameter is defined as the
nonequilibrium state. As the value of the control parameter
increases, the system recedes away from equilibrium state. By Gibbs'
theorem, we know that mere substitution of these distributions into
BG entropy will yield nonsensical results due to the failure of the
equal mean energy assumption because of the interaction with the
environment. Therefore, Klimontovich [12-17] renormalizes the
equilibium distribution i.e., $r_{eq}$. However, we note that the
S-theorem of Klimontovich is not limited to the comparisons of
equilibrium and nonequilibrium distributions. It can be used to
compare any two stationary distributions as long as they are
different from one another through the change in the value of the
control parameter. From now on, we will drop the subscript $(eq)$
from the equilibrium probability distribution $r$. Therefore, it
should be understood that the probability distribution $r$ denotes
the equilibrium distribution. Next, Klimontovich defines a quantity
called effective energy $U_{\text{eff}}$ [12-17] in terms of the
equilibrium state as

\begin{equation}\label{ee}
U_{\text{eff}}=-\ln r.
\end{equation}

\noindent The definition of effective energy in Eq. (\ref{ee}) is
central to the S-theorem, and therefore requires some explanation.
The application of the effective energy definition in Eq. (\ref{ee})
to the equilibrium distribution results in
$U_{\text{eff}}=\beta\varepsilon_{i}$ (apart from normalization).
This explains why it is called effective energy since it is
proportional to the multiplication of the Lagrange multiplier
$\beta$ associated with the internal energy constraint and the
energy of the $i$th microstate. Consequently, its average
corresponds to the $\beta$ times the average internal energy of the
equilibrium distribution. We then equate the effective energy of the
equilibrium and nonequilibrium states and this equalization
procedure is called \textit{renormalization} by Klimontovich. As a
result of this renormalization, we no longer use the equilibrium
distribution $r$, but the renormalized equilibrium distribution
$\widetilde{r}$, where tilde denotes that the renormalization is
taken care of. Explicitly written, the renormalization reads

\begin{equation}\label{renormalization1}
\left\langle U_{\text{eff}}\right\rangle ^{(req)}=\left\langle
U_{\text{eff}}\right\rangle ^{(neq)},
\end{equation}

\noindent where superscripts $(req)$ and $(neq)$ denote the
renormalized equilibrium and ordinary nonequilibrium states,
respectively. Then, S-theorem is equivalent to showing that the
renormalized entropy $R[p\Vert{\widetilde{r}}]$ defined as

\begin{equation}\label{sthm}
R[p\Vert{\widetilde{r}}]\equiv\
S_{BG}^{neq}(p)-\widetilde{S}_{BG}^{eq}(\widetilde{r})
\end{equation}

\noindent is negative i.e., $R[p\Vert{\widetilde{r}}]<0$, since this
implies that $\widetilde{S}_{BG}^{eq}>S_{BG}^{neq}$, where $S_{BG}$
is the usual BG entropy given by

\begin{equation}\label{bg}
S_{BG}(p)=-\sum_{i=1}^{W}p_{i}\ln p_{i},
\end{equation}

\noindent where $p_{\text{i }}$ is the probability of the system in
the $i$th microstate, $W$ is the total number of the configurations
of the system.

The proof of the S-theorem relies on the definition of
Kullback-Leibler relative entropy (KL) [26], which reads

\begin{equation}\label{KL}
K[p\Vert{r}]=\sum_{i=1}^{W}p_{i}\ln \left(\frac{p_{i}}{r_{i}}\right),
\end{equation}

\noindent since the renormalized entropy given in Eq. (\ref{sthm})
is, through the renormalization in Eq. (\ref{ee}), equal to

\begin{equation}\label{rkl}
R[p\Vert{\widetilde{r}}]=-K[p\Vert{\widetilde{r}}].
\end{equation}

\noindent The S-theorem is therefore proven [27], since KL entropy
is positive definite i.e.,

\begin{equation}\label{proof1}
R[p\Vert{\widetilde{r}}]<0.
\end{equation}

\noindent Although Klimontovich's S-theorem is a generalization of
the Gibbs' theorem, it is based on the use of exponential stationary
distributions and therefore BG entropy. Therefore, the S-theorem due
to Klimontovich cannot be used in the case of inverse power law
stationary distributions. In order to overcome this difficulty, a
new generalization of S-theorem has been given in the framework of
nonadditive thermostatistics. This nonadditive S-theorem first
generalizes the effective energy by deforming the logarithm in Eq.
(\ref{ee}) as

\begin{equation}\label{ee2}
U_{\text{eff}}=\ln _{q}\left(\frac{1}{r_{eq}}\right),
\end{equation}

where the $q$-logarithm function $\ln _{q}(x)$ is defined as

\begin{equation}\label{qlog}
\ln _{q}(x)=\frac{x^{1-q}-1}{1-q}.
\end{equation}

The nonadditive S-theorem can be similarly written as

\begin{equation}\label{rq}
R_{q}[p\Vert\widetilde{r}]=S_{q}(p)-\widetilde{S}_{q}(\widetilde{r})=-K_{q}[p\Vert
\widetilde{r}]<0,
\end{equation}

\noindent for positive values of the nonadditivity parameter $q$,
where the nonadditive Tsallis entropy $S_{q}$ [24, 25] is given by

\begin{equation}\label{tsallis}
S_{q}(p)=\frac{\sum_{i}^{W}p_{i}^{q}-1}{1-q}
\end{equation}

\noindent and the nonadditive relative entropy $K_{q}[p\Vert
\widetilde{r}]$ [28] is given by

\begin{equation}\label{kq}
K_{q}[p\Vert \widetilde{r}]=\frac{\sum_{i}p_{i}^{q}}{q-1}+\sum_{i}\widetilde{r}%
_{i}^{q}-
\frac{q}{q-1}\sum_{i}p_{i}\widetilde{r}_{i}^{q-1} .
\end{equation}

\noindent The proof of the nonadditive S-theorem is limited to the
positive $q$ values only, since the nonadditive relative entropy is
positive only in this region. However, we can exclude negative $q$
values, since Tsallis entropy is stable only for positive values of
$q$ [29].

It should be noted that the ordinary S-theorem by Klimontovich is
recovered in the $q\rightarrow1$ limit. This can be seen from the
inspection of Eq. (\ref{rq}), since Tsallis entropy expressions
become BG entropies, whereas all the stationary distributions become
exponential in the $q\rightarrow1$ limit. The nonadditive relative
entropy in Eq. (\ref{kq}) becomes KL relative entropy in the
aforementioned limit i.e., $K[p\Vert r]\equiv \sum_{i}p_{i}\ln
(p_{i}/r_{i})$ [26, 28], which is positive definite, ensuring the
negativity of the ordinary renormalized entropy.

\section{optical lattices and Modified S-theorem}

In the previous section, we reviewed two interrelated criteria of
self-organization, namely, the additive and nonadditive S-theorems.
The former can be used when there are exponential stationary
distributions so that one can use BG entropy together with the
renormalization of effective energies of stationary states. The
latter, on the other hand, is used when the stationary distributions
are $q$-exponentials associated with the Tsallis entropy. However,
self-organization is not limited to the transitions amongst
exponential or $q$-exponential stationary distributions. An
intermediate transition is possible so that a transition from an
exponential to $q$-exponential stationary distribution may be
observed. In this case, neither additive nor nonadditive S-theorems
can be used.

Such a transition has recently been observed in anomalous transport
experiments in optical lattices [10]. Before proceeding further, it
is important to review how the transition from an ordinary
exponential stationary distribution to the $q$-exponential
distribution can occur [9]. In order to do so, one first needs to
derive the master equation, which describes the atom-laser
interaction in the optical lattice. Assuming that the laser
intensity is low and the atoms move fast, a semiclassical treatment
is possible, which allows one to obtain, after spatial averaging,
so-called Rayleigh equation for the semiclassical Wigner function
$W(p,t)$

\begin{equation}\label{wigner}
\frac{\partial W}{\partial t}=-\frac{\partial }{\partial p}\left[ K(p)W%
\right] +\frac{\partial }{\partial p}\left[ D(p)\frac{\partial W}{\partial p}%
\right],
\end{equation}

\noindent where the variables $p$ and $t$ denote momentum and time,
respectively [30, 31, 32]. The drift coefficient $K(p)$ and
diffusion factor $D(p)$ satisfy the relation [33]

\begin{equation}\label{drift}
\frac{K\left( p\right) }{D\left( p\right) }=-\frac{\beta }{1-\beta
\left( 1-q\right) U\left( p\right) }\frac{\partial U\left( p\right)
}{\partial p},
\end{equation}

\noindent where

\begin{equation}\label{control}
q=1+a_{c}\;\;\;\;\; ; \;\;\;\;\; U\left( p\right) =p^{2}.
\end{equation}

\noindent The parameter $\beta$ is positive and can be expressed in
terms of the microscopic damping and diffusion coefficients [9]. The
term $a_{c}$ denotes the control parameter, which represents the
interaction of the atom with the environment i.e., laser field. The
control parameter can be written more explicitly in terms of the
microscopic parameters [9] as

\begin{equation}\label{control2}
a_{c}=\frac{2D_{0}}{\alpha p_{c}^{2}}.
\end{equation}

\noindent The terms $\alpha$ and $p_{c}$ denote the damping
coefficient and the capture momentum, respectively. The term $D_{0}$
is the diffusion term associated with the fluctuations due to
spontaneous photon emissions and fluctuations in the difference of
photons absorbed in the two laser beams [9]. Note that the control
parameter $a_{c}$ can also be written in terms of the potential
depth $U_{0}$ and the recoil energy $E_{R}$ [30] as

\begin{equation}\label{control3}
a_{c}=\frac{44E_{R}}{U_{0}}.
\end{equation}

\noindent Since the condition (\ref{drift}) is satisfied in optical
lattices, the exact stationary solution to the Rayleigh equation in
Eq. (\ref{wigner}) is given in terms of the $q$-Gaussians

\begin{equation}\label{q-gaussian}
W_{q}\left( p\right) =C_{q}\left[ 1-\beta \left( 1-q\right) U\left( p\right) %
\right] ^{1/\left( 1-q\right) },
\end{equation}

\noindent where we also require $W_{q}\left( p\right) \rightarrow 0$
when $p\rightarrow \pm \infty $  When the control parameter is equal
to zero, the nonadditivity parameter $q$ becomes equal to 1. We then
have stationary equilibrium distribution of the Gaussian form i.e.,

\begin{equation}\label{exp}
W_{1}\left( p\right) =C_{1}\exp \left[ -\beta U\left( p\right)
\right].
\end{equation}

\noindent As the interaction between the atoms and the laser field
is turned on, the control parameter $a_{c}$ takes nonzero positive
values. As a result, the value of the nonadditivity parameter $q$
exceeds 1 so that we have stationary nonequilibrium distribution of
$q$-Gaussian form given by Eq. (\ref{q-gaussian}).

As we remarked earlier, this transition from an ordinary exponential
stationary distribution to a $q$-Gaussian distribution cannot be
analyzed within the frameworks of the aforementioned versions of the
S-theorem. Therefore, we need a modified S-theorem, which can be
used as a criterion of self-organization for the transition from
Gaussian to $q$-Gaussian regime.

Since the equilibrium distribution is an ordinary Gaussian, the
modified S-theorem must include BG entropy associated with the
Gaussian equilibrium distribution as is the case with the original,
additive S-theorem by Klimontovich [12-17]. On the other hand, the
stationary nonequilibrium distribution is a $q$-Gaussian. Therefore,
the entropy of the stationary nonequilibrium distribution must be
calculated by using Tsallis entropy as in the nonadditive S-theorem
[24, 25]. The modified S-theorem will be the difference of these two
entropies together with the renormalization condition.

The only ingredient left is how the renormalization will be carried
out in the modified S-theorem. However, the renormalization
condition can be borrowed from Klimontovich's S-theorem  [12-17] as
it is, since the equilibrium distribution associated with the
original and modified S-theorems both is of the same form i.e.,
exponential. Therefore, similar to Eq. (\ref{ee}), we can write

\begin{equation}\label{ee3}
U_{\text{eff}}=-\ln W_{1}\left( p\right)
\end{equation}

\noindent and then use Eq. (\ref{renormalization1}) to obtain

\begin{equation}\label{renormalization2}
\left\langle p^{2}\right\rangle ^{(req)}=\left\langle
p^{2}\right\rangle ^{(neq)},
\end{equation}

\noindent where we have inserted Eq. (\ref{ee3}). The renormalized
entropy $\Re[W_{q}\Vert\widetilde{W}_{1}]$ corresponding to the
modified S-theorem can then be written as

\begin{equation}\label{modifieds}
\Re[W_{q}\Vert\widetilde{W}_{1}]=S_{q}(W_{q})-\widetilde{S}_{BG}(\widetilde{W}_{1}).
\end{equation}

\noindent Similar to the additive and nonadditive versions of
S-theorem, we calculate $\Re[W_{q}\Vert\widetilde{W}_{1}]$ and see
whether it becomes negative for some interval of $q$ values. Before
proceeding with the renormalization condition given by Eq.
(\ref{renormalization2}), we write the equilibrium

\begin{equation}\label{w1}
W_{1}\left( p\right) =\sqrt{\frac{4\beta }{\pi }}\exp \left[ -\beta p^{2}%
\right]
\end{equation}

\noindent and the stationary nonequilibrium distribution

\begin{equation}\label{wq}
W_{q}\left( p\right) =\sqrt{4\beta \left( q-1\right) }\left[ B\left( \frac{1%
}{2},\frac{3-q}{2q-2}\right) \right] ^{-1}\left[ 1-\beta \left(
1-q\right) p^{2}\right] ^{1/\left( 1-q\right) }\;\;\; ; \;\;\; 1<q<3,
\end{equation}

\noindent by inserting the normalization constants $C_{1}$ and
$C_{q}$ explicitly. The term $B \left( x,y\right) $ denotes the Beta
function [34]. We now rewrite Eq. (\ref{renormalization2})

\begin{equation}\label{renormalization3}
\int\limits_{0}^{\infty }dpp^{2}\widetilde{W}_{1}\left( p\right)
=\int\limits_{0}^{\infty }dpp^{2}W_{q}\left( p\right).
\end{equation}

\noindent By substituting Eqs. (\ref{w1}) and (\ref{wq}) into Eq.
(\ref{renormalization3}), we obtain the renormalized inverse
temperature $\widetilde{\beta}$ as

\begin{equation}\label{renormalizedbeta}
\widetilde{\beta }=\frac{\left( q-1\right) \Gamma \left( \frac{3-q}{2q-2}%
\right) }{\Gamma \left( \frac{5-3q}{2q-2}\right)
}\beta \;\;\; ; \;\;\; 1<q<\frac{5}{3},
\end{equation}

\noindent where $\Gamma \left( x,y\right) $ is the Gamma function
[34]. The associated equilibrium

\begin{equation}\label{wwbg}
\widetilde{S}_{BG}\left( \widetilde{W}_{1}\right) =\frac{1}{2}-\frac{1}{2}%
\ln \left( \frac{4\widetilde{\beta }}{\pi }\right)\;\;\; ; \;\;\; 1<q<\frac{5}{3}
\end{equation}

\noindent and nonequilibrium entropies can be calculated

\begin{equation}\label{wwqq}
S_{q}\left( W_{q}\right) =\frac{1-\left[ 4\beta \left( q-1\right) \right] ^{%
\frac{q-1}{2}}\left[ B\left( \frac{1}{2},\frac{3-q}{2q-2}\right)
\right] ^{-q}B\left( \frac{1}{2},\frac{q+1}{2q-2}\right)
}{q-1}\;\;\; ; \;\;\; 1<q<\frac{5}{3},
\end{equation}

\noindent where the renormalized inverse temperature
$\widetilde{\beta}$ in Eq. (\ref{wwbg}) can be substituted from Eq.
(\ref{renormalizedbeta}). Finally, the renormalized entropy reads

\begin{eqnarray}\label{final}
\Re[W_{q}\Vert\widetilde{W}_{1}] =\frac{1-\left[ 4\beta \left( q-1\right) \right] ^{%
\frac{q-1}{2}}\left[ B\left( \frac{1}{2},\frac{3-q}{2q-2}\right)
\right]
^{-q}B\left( \frac{1}{2},\frac{q+1}{2q-2}\right) }{q-1}-\nonumber \\
\frac{1}{2}+\frac{1}{%
2}\ln \left[ \frac{4\left( q-1\right) \Gamma \left(
\frac{3-q}{2q-2}\right) }{\Gamma \left( \frac{5-3q}{2q-2}\right) \pi
}\beta \right]\;\;\; ; \;\;\; 1<q<\frac{5}{3}.
\end{eqnarray}

\noindent The renormalized entropy $\Re[W_{q}\Vert\widetilde{W}_{1}]
$ in Eq. (\ref{final}) is indeed negative in the interval
$1<q<\frac{5}{3}$ as can be seen from Fig. 1, independent of the
value of $\beta$. It corresponds to an entropy decrease when the
entropies of the states are evaluated with the consistent
expressions of entropy i.e., the Gaussian distribution at
equilibrium with BG entropy and $q$-Gaussian distributions at
nonequilibrium with the Tsallis entropy. As the control parameter
$a_{c}$ increases from zero (i.e., as the nonadditivity parameter
$q$ increases from 1), the system moves from Gaussian distribution
to $q$-Gaussian distributions, decreasing its entropy. This
indicates that the transition from the Gaussian distribution to the
$q$-Gaussian distributions in dissipative optical lattices is
self-organization, signifying a transition to a more ordered state.
It can be seen from Fig. 2b of Douglas \textit{et al.} [10] that the 
self-organization interval determined by the modified S-theorem i.e., 
$1<q<\frac{5}{3}$, is confirmed by the experiment, although the whole 
range of admissible $q$ values is theoretically $1<q<3$.

We also remark that the renormalized inverse temperature
$\widetilde{\beta}$ is a linear function of the ordinary inverse
temperature $\beta$ as can be seen from 
Eq.(\ref{renormalizedbeta}). Therefore, the procedure of
renormalization does not deform the equilibrium distribution, which
is an exponential. Its role is to compensate the difference between
the equilibrium temperature associated with the Gaussian
distribution and the parameter $\beta$ in the $q$-Gaussian
distribution, since the latter is not physical temperature. This
compensation is equivalent to heating up the equilibrium temperature
of the system as a result of the linear decrease of
$\widetilde{\beta}$ with respect to $\beta$ for all $q$ values in
the range $1<q<\frac{5}{3}$. The fact that $\widetilde{\beta}$
depends explicitly on $q$ implies that the renormalization also
takes the microscopic structure into account.

We finally mention the effect of renormalization on the spatial
correlation function $G\left( a\right) =\int dx\psi \left(
x,t\right) \psi ^{\ast }\left( x+a,t\right)$, where $\psi \left(
x,t\right)$ denotes the wave function at time $t$ [35]. The spatial
correlation function can be considered as a measure of how coherent
a state is between two different points. For a Gaussian wave packet
such as $W_{1}$, the spatial correlation function too is a Gaussian
with a correlation length $\lambda =2 \hslash \sqrt{\beta }$ [9].
The renormalized correlation length, on the other hand, is equal to
$\widetilde{\lambda} =2 \hslash 
\sqrt{\left(q-1\right)\Gamma\left(\frac{3-q}{2q-2}\right) / 
\Gamma\left(\frac{5-3q}{2q-2}\right)\beta}$ in the
interval $1<q<\frac{5}{3}$. Therefore, we conclude that the
correlation length decreases with increasing value of the
nonadditivity parameter $q$. In other words, the Gaussian
distribution becomes less coherent spatially as a result of the
renormalization procedure employed in the modified S-theorem. This
decrease in coherence becomes more dominant as the system moves
further from equilibrium.

\section{Conclusions}

By analyzing the intermediate anomalous transport regime in
dissipative optical lattices [9, 10] in the framework of a modified
version of Klimontovich's S-theorem, we have shown that the
self-organization occurs in the transition from Gaussian to the
$q$-Gaussian distributions, implying an entropy decrease i.e., a
more ordered state as the system is driven out of equilibrium as a
result of the interaction of the atoms with the laser field.

An interesting result of the modified S-theorem is that it confines
the self-organization in the region $1<q<\frac{5}{3}$, although, 
for example, there is nothing which forbids transition to a region with $q=2$, 
(note that the $q$-Gaussian distributions are normalizable in the whole 
range $1<q<3$). This is also confirmed with the experimental results 
obtained by Douglas \textit{et al.} [10]. 
The reason for this confinement is due to the
additional requirement associated with the (modified) S-theorem
i.e., renormalization of the effective energy. For dissipative
optical lattices, the effective energy term corresponds to the
square of the momentum. Therefore, the procedure of renormalization
confines the self-organization to the region, where the second
moment of the distribution $\left\langle p^{2}\right\rangle $
converges. This is similar to the requirement of finite capture
momentum for the occurrence of $q$-Gaussian distribution in
dissipative optical lattices [9]. If the capture momentum $p_{c}$ is
infinite, then the control parameter $a_{c}$ becomes zero according
to Eq. (\ref{control2}). Since zero value of the control parameter
corresponds to the case $q=1$, we obtain only Gaussian dynamics,
eliminating the possibility of an anomalous transition [9]. The
existence of the non-Gaussian transition requires both finite
capture momentum $p_{c}$ and finite second moment $\left\langle
p^{2}\right\rangle $.

We also conclude that the correlation length associated with the
equilibrium distribution decreases with increasing value of the
nonadditivity parameter $q$ as a result of the renormalization
procedure. This implies that the equilibrium Gaussian distribution
becomes less coherent spatially as the system moves further from
equilibrium.

We finally note that another version of the modified S-theorem was
provided by Bashkirov and Vityazev in Ref. [36] based on R\'{e}nyi
entropy [37]. However, the compared distributions were both chosen
as the inverse power law distributions although the renormalized
entropy in Ref. [36] was defined as the difference of R\'{e}nyi and
BG entropies.

\section{ACKNOWLEDGEMENTS}
This work has been supported by TUBITAK (Turkish Agency) under the
Research Project number 108T013.

\newpage

\begin{figure}
 \begin{center}
\includegraphics[scale = 0.8]{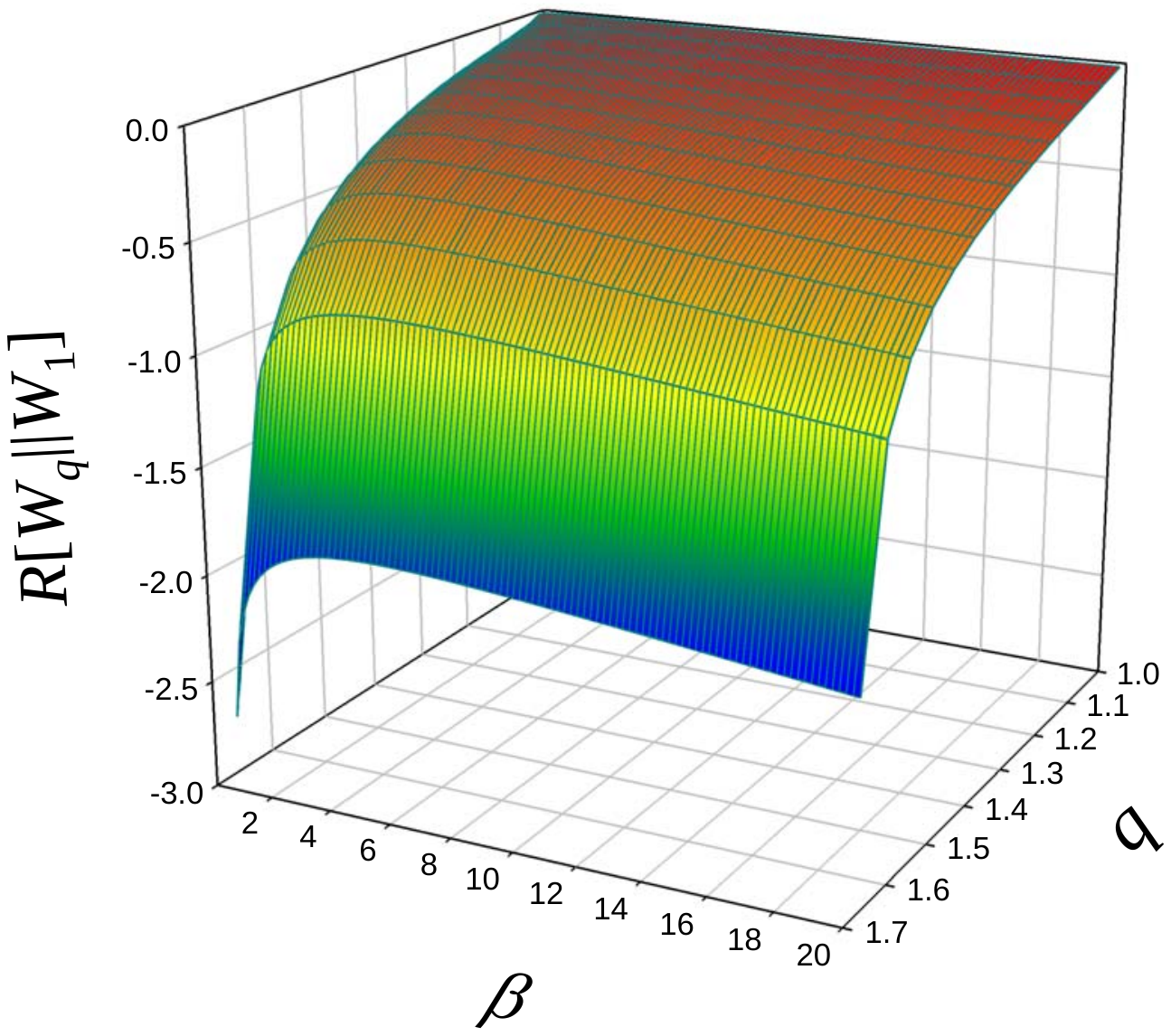}
\caption{The renormalized entropy $\Re[W_{q}\Vert\widetilde{W}_{1}]$ 
in Eq. (\ref{final}) is plotted in the interval $1<q<\frac{5}{3}$ for 
a representative range of $\beta$, from which the negativity of the 
renormalized entropy is evident.}
 \end{center}
\end{figure}

\end{document}